\documentclass[%
aip,
amsmath,amssymb,
reprint,%
]{revtex4-1}

\usepackage{siunitx}
\usepackage{todonotes}
\usepackage{csquotes}

\usepackage{tabularx}
\newcolumntype{Y}{>{\centering\arraybackslash}X}
\usepackage{threeparttable}

\usepackage{hyperref}
\usepackage{cleveref}

\usepackage{colortbl}
\normalsize
\usepackage{booktabs}

\crefformat{figure}{FIG.~#2#1#3}
\crefformat{equation}{Eq.~(#2#1#3)}

\begin{document}
	
\title{A tunable fiber Fabry-Perot cavity for hybrid optomechanics stabilized at 4~K}
	
\date{\today}
	
\author{Thibaud Ruelle}
\affiliation{Department of Physics, University of Basel, 4056 Basel, Switzerland}

\author{David Jaeger}
\affiliation{Department of Physics, University of Basel, 4056 Basel, Switzerland}
\affiliation{Swiss Nanoscience Institute, University of Basel, 4056 Basel, Switzerland}

\author{Francesco Fogliano} 
\affiliation{Department of Physics,  University of Basel, 4056 Basel, Switzerland}

\author{Floris Braakman}
\affiliation{Department of Physics, University of Basel, 4056 Basel, Switzerland}

\author{Martino Poggio}
\affiliation{Department of Physics, University of Basel, 4056 Basel, Switzerland}
\affiliation{Swiss Nanoscience Institute, University of Basel, 4056 Basel, Switzerland}
\email{martino.poggio@unibas.ch}

\begin{abstract}
  We describe an apparatus for the implementation of hybrid optomechanical systems at \SI{4}{\kelvin}.
  The platform is based on a high-finesse, micrometer-scale fiber Fabry-Perot cavity, which can be widely tuned using piezoelectric positioners.
  A mechanical resonator can be positioned within the cavity in the object-in-the-middle configuration by a second set of positioners.
  A high level of stability is achieved without sacrificing either performance or tunability, through the combination of a stiff mechanical design, passive vibration isolation, and an active Pound-Drever-Hall feedback lock incorporating a reconfigurable digital filter.
  The stability of the cavity length is demonstrated to be better than a few picometers over many hours both at room temperature and at \SI{4}{\kelvin}.
 \end{abstract}

\maketitle

\section{Introduction}

Driven by improvements in the fabrication and control of nanomechanical resonators \cite{PootMechanicalsystemsquantum2012}, researchers have proposed enhancing the capabilities of optomechanical systems by combining cavity optomechanics \cite{AspelmeyerCavityoptomechanics2014} (COM) with cavity quantumelectrodynamics (CQED) \cite{KimbleStronginteractionssingle1998} within a single platform.  
Such hybrid optomechanical systems \cite{KurizkiQuantumtechnologieshybrid2015,Chuperspectivehybridquantum2020a} promise increased optomechanical coupling \cite{HammererStrongCouplingMechanical2009}, quantum control of mechanical resonators \cite{BergholmOptimalcontrolhybrid2019a}, quantum technology applications \cite{KurizkiQuantumtechnologieshybrid2015}, and access to the physics of a fully coupled tripartite system \cite{RestrepoFullycoupledhybrid2017}.  
Hybrid optomechanical systems have already been implemented in some platforms, including trapped atomic gases \cite{CamererRealizationOptomechanicalInterface2011}, microwave circuit devices \cite{PirkkalainenHybridcircuitcavity2013}, and optomechanical crystals \cite{TianAllopticaldynamicmodulation2022}.

Due to their open nature, Fabry-Perot cavities are ideal for the realization of hybrid optomechanical systems via a mechanical resonator \cite{TreutleinHybridMechanicalSystems2014} in an object-in-the-middle configuration \cite{FaveroMechanicalResonatorsMiddle2014}.  
Such hybrid systems are particularly flexible, because they are both spectrally and spatially tunable and they allow for the study of a wide variety of mechanical resonators.  
However, conflicting requirements for high finesse and small mode volume make their implementation challenging \cite{CernotikInterferenceeffectshybrid2019}.

Because of their small size, fiber Fabry-Perot cavities (FFPCs) \cite{HungerfiberFabryPerot2010} are the natural choice to satisfy
these demands, leading to their successful use in state-of-the-art CQED \cite{GallegoStrongPurcellEffect2018,TakahashiStrongCouplingSingle2020} and COM \cite{RochauDynamicalBackactionUltrahighFinesse2021,FoglianoMappingCavityOptomechanical2021} systems.  
The use of FFPCs, however, is limited by their sensitivity to mechanical noise, especially in cryogenic environments.  
Although tremendous effort has been dedicated to improving their stability \cite{Zhongmillikelvinallfibercavity2017,SalzCryogenicplatformcoupling2020,JanitzHighmechanicalbandwidth2017,Fontanamechanicallystabletunable2021a,VadiaOpenCavityClosedCycleCryostat2021}, most implementations make sacrifices in terms of either finesse, tunability, or cryo-compatibility.

Here, we present an experimental platform designed for the implementation of hybrid optomechanical systems in the object-in-the-middle configuration. 
The apparatus is built around a high-finesse, micrometer-scale FFPC, which can be widely tuned using piezoelectric positioners.
The combination of a stiff mechanical design, passive vibration isolation, and active stabilization using the Pound-Drever-Hall (PDH) technique combined with a digital filter, results in a mechanical stability of a few picometers over many hours both at room temperature and at \SI{4}{\kelvin}.

\begin{figure*}[htb]
	\centering\includegraphics[width=17cm]{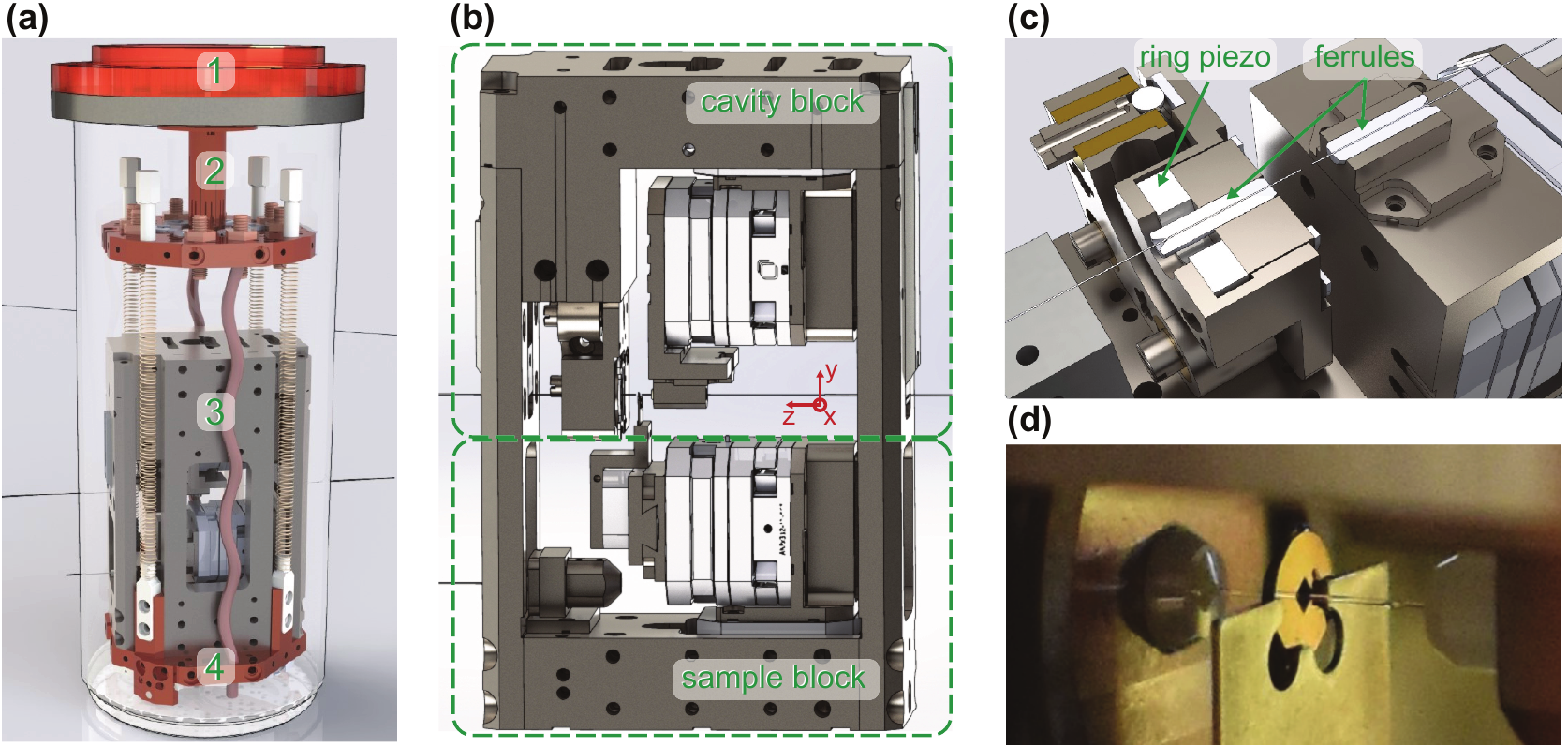}
	\caption{
		(a): Overview of the bottom of the cryogenic probe with (1) the copper plate at the top of the can, which is in direct contact with the helium bath, and (2) a second copper plate, from which the cavity apparatus is suspended.
		All optical fibers, electrical connections, and the copper braids used to thermalize the cavity are thermally connected to (2). (3) The titanium frame and (4) copper base plate surrounding the optical cavity are suspended from (2) via copper-beryllium springs.
		(b): The titanium frame containing the cavity block, which is used to align and tune the FFPC, and the sample block, which is used to position a mechanical resonator inside the FFPC.
		(c): Cutout view of the cavity block.
		(d): Photograph of a membrane held in the middle of the FFPC within the assembled probe. 
	}
	\label{fig:platform_edit}
\end{figure*}

\section{The experimental platform}

The experimental platform includes the cryogenic probe, which hosts the fiber-based cavity optomechanical system, and the laser setup used to address it.

\subsection{The cryogenic probe}
The cryogenic probe is designed to be inserted into a liquid-helium bath cryostat (Cryomagnetics), which includes a superconducting solenoid capable of producing \SI{8}{\tesla} of magnetic field and a liquid nitrogen shield to minimize the helium boil-off rate.
This cryostat is placed with additional weights on top of a passive vibration isolation platform (Minus K 1000BM-1CMM) with a cut-off frequency of a few \si{\hertz}, which provides shielding from seismic noise.

The optomechanical system is hosted at high-vacuum in a \SI{10}{\cm}-diameter and \SI{25}{\cm}-long cylindrical can near the bottom of the probe, shown in \cref{fig:platform_edit}(a).  
Once the probe is inserted into the cryostat, the can containing the experimental apparatus is centered in the bore of the superconducting magnet.
A copper plate at the top of the can, labeled as (1) in \cref{fig:platform_edit}(a), is in direct contact with the helium bath.
All optical fibers, electrical connections, and the copper braids are fed through and thermalized on a second copper plate, labeled (2) in \cref{fig:platform_edit}(a), which is directly bolted to (1).
This second plate also supports the structure surrounding the optical cavity, which consists of a titanium frame and a copper base-plate, labeled (3) and (4), respectively.
This structure is suspended from (2) by four \SI{16}{\cm}-long copper-beryllium springs with a spring constant of \SI{0.023}{\newton\per\mm}, which isolate it from external mechanical vibrations.
Two soft copper braids, arranged so as to minimally influence the suspension system, provide a thermal link between the suspended cavity and (2).
A sensor mounted on (4) is used to monitor the temperature of the cavity.
The probe reaches pressures lower than a few \SI{d-7}{\milli\bar} at room temperature using turbo and ion pumps and a few \SI{d-8}{\milli\bar} when cryopumping at \SI{4}{\K}.

The optical cavity itself and the mechanical resonator are mounted within the titanium frame pictured in \cref{fig:platform_edit}(b).
The frame contains two blocks, each holding one part of the optomechanical system: the cavity block and the sample block.
The design of the titanium frame, which includes side plates (not shown in \cref{fig:platform_edit} (b) for clarity, but visible in (a)), is optimized for mechanical rigidity, stability, and to minimize misalignments due to thermal contraction.

The cavity block holds the FFPC. 
As depicted in \cref{fig:platform_edit}(c), a fiber mirror is held in a fixed mount attached to a stack of 3 piezoelectric positioners (Attocube ANPx312), providing \SI{6}{\mm} of linear slip-stick motion in the $x$, $y$ and $z$-directions.
On the left side of \cref{fig:platform_edit}(c), a second fiber mirror is held in a titanium disk inside the faceplate of a custom-made kinematic mount.
As visible in the cutout shown in \cref{fig:platform_edit}(c), the titanium disk is clamped against a piezoelectric ring by a leaf spring.
The kinematic mount allows for manual tip and tilt adjustment of the cavity alignment.
Unlike the $x$, $y$, and $z$ positions, which can be modified with the system closed and at low temperature, the tip and tilt can only be adjusted while the system is open.
The piezoelectric ring provides short-range, high-bandwidth (up to \SI{100}{\kHz}) tunability useful for the stabilization and fast scanning of the cavity length.

The FFPC is formed by two fiber mirrors which are machined using a CO\textsubscript{2} laser ablation technique \cite{HungerLasermicrofabricationconcave2012}.  
Ruelle et al.~\cite{RuelleOptimizedsingleshotlaser2019} describes the ablation procedure and the fitting procedure used to extract the geometrical characteristics of the fiber mirrors, e.g.\ their radius of curvature $\mathrm{ROC}$, depth $t$, and spherical diameter $D_\mathrm{sph}$.
The FFPC for measurements shown in this article is formed between a single-mode fiber (SMF) (IVG fiber Cu800) and a multi-mode fiber (MMF) (IVG fiber Cu50/125).
The geometry of the fiber-mirror at the end of the SMF is: $\mathrm{ROC} = \SI{92}{\um}$, $t = \SI{0.2}{\um}$, and $D_\mathrm{sph} = \SI{14}{\um}$; while the geometry at the end of the MMF is: $\mathrm{ROC} = \SI{57}{\um}$, $t = \SI{1.1}{\um}$, and $D_\mathrm{sph} = \SI{19}{\um}$.

The sample block holds the mechanical resonator and is used to position it within the FFPC.
This block employs an identical stack of piezoelectric positioners as the cavity block, allowing for adjustment of the resonator's $x$, $y$, and $z$ position with the system either open or closed and at low temperature. 
Furthermore, the sample holder is fixed to a piezoelectric $xy$-scanner (Attocube ANSxy50) providing 30x\SI{30}{\um} (15x\SI{15}{\um}) of continuous motion at room temperature (\SI{4}{\kelvin}).

\begin{figure*}[htb]
	\centering\includegraphics[width=17cm]{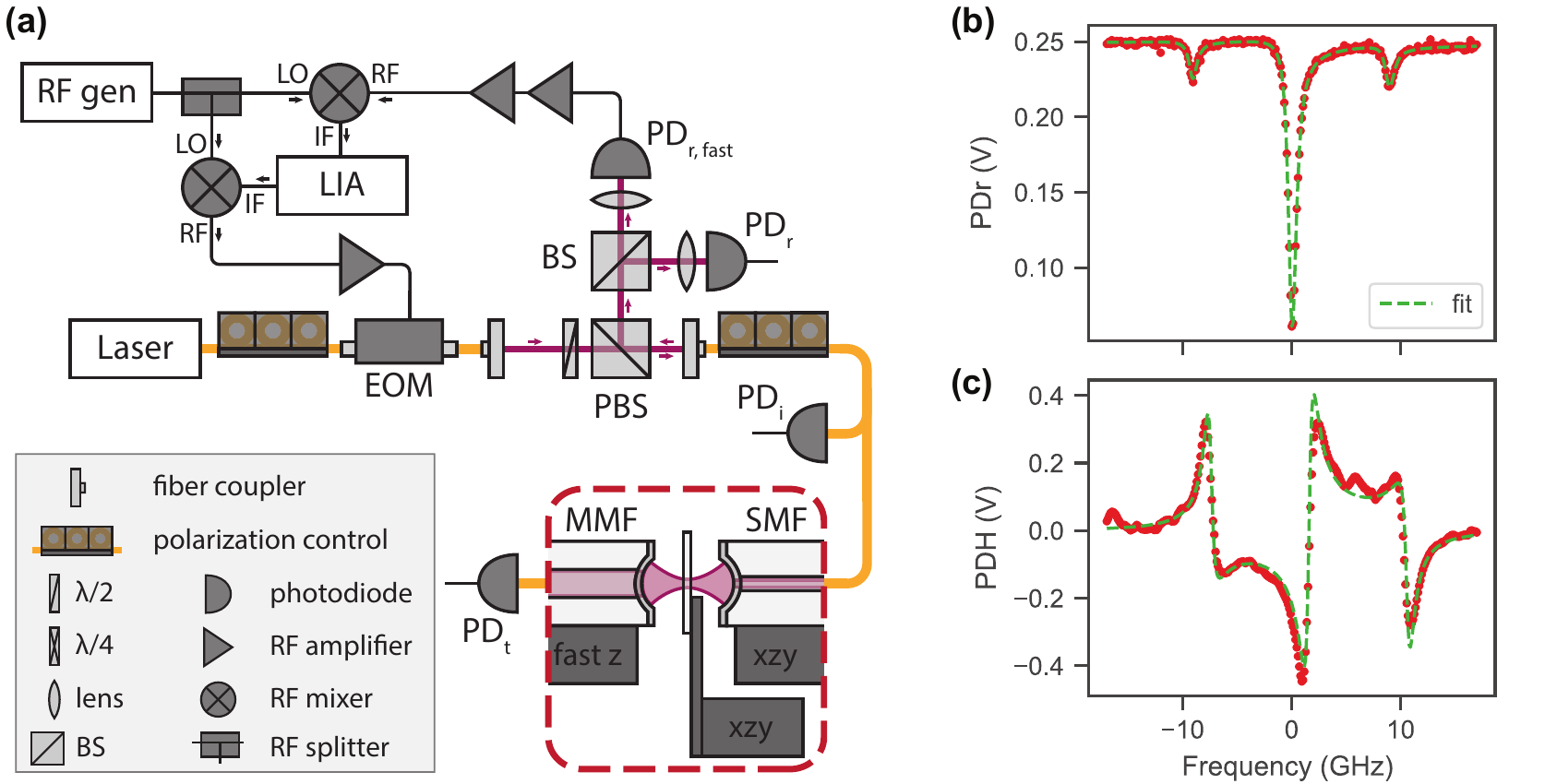}
	\caption{ 
		(a) Overview of the laser setup, including its connection to the FFPC (within red dashed box).
		Measured photodiode signals include the slow and fast reflected signals $\text{PD}_{\text{r}}$ and $\text{PD}_{\text{r,fast}}$, respectively, the transmitted signal $\text{PD}_{\text{t}}$, and the intensity monitor $\text{PD}_{\text{i}}$.
		LIA: lock-in amplifier, EOM: electro-optic modulator, PD: photodiode, PBS/BS: (polarizing) beamsplitter, SMF/MMF: single/multi-mode fiber.
		(b) Measured signal reflected from the FFPC while scanning the cavity length over a small range, revealing the main cavity resonance in the middle and two sidebands generated by the phase modulation introduced by the EOM.
		(c) The resulting Pound-Drever-Hall error signal revealing a linear region around the resonance position.
	}
	\label{fig:setup}
\end{figure*}

\subsection{The laser setup}
\cref{fig:setup} (a) depicts the external laser setup, designed to interface with the cryogenic probe.
This setup is used to prepare the laser beam before sending it to the cavity, as well as for the collection and processing of the reflected and transmitted beams returning from the system.
Optical signals detected here are used for various types of measurements, as well as for the active stabilization of the cavity length.

A wavelength-tunable external-cavity diode laser (Toptica CTL950) (\SI{915}{\nm} to \SI{985}{\nm}) serves as the source of the signal.
Its phase is modulated using a fiber-coupled electro-optic modulator (EOM) (iXblue NIR-MPX950-LN-10), which is placed after a fiber-based polarization controller.
The radio-frequency (RF) signal sent to the EOM has an angular frequency $\omega_\mathrm{RF} = \omega_\mathrm{LO} + \omega_\mathrm{IF}$ and, as shown in \cref{fig:setup} (a), is generated by mixing a local oscillator (LO) from an RF source (Rohde \& Schwarz SGS100A) and the reference at intermediate frequency (IF) of a lock-in amplifier (LIA) (Zurich Instruments HF2LI).
This RF signal is amplified (iXblue DR-AN-10-HO) before entering the EOM.

After passing through the EOM, the light enters a free-space optical setup, in which the signal is sent to the FFPC through a polarizing beamsplitter. 
The intensity of the laser light is monitored by a variable gain photoreceiver (FEMTO OE-300), $\text{PD}_{\text{i}}$.
Light transmitted through the FFPC is measured by a second variable gain photoreceiver (FEMTO OE-300), $\text{PD}_{\text{t}}$, while reflected light propagates back up the optical fiber.
The reflected signal is collected using a polarization-based dark-field technique \cite{Kuhlmanndarkfieldmicroscopebackgroundfree2013}, preventing light reflected from the cavity from being fed back into the laser and allowing for collection with near-unity efficiency.
The reflected light is split onto two photodiodes: a third variable gain photoreceiver (FEMTO OE-300) to measure the slow part of the signal, $\text{PD}_{\text{r}}$, and a photodiode with a bandwidth of \SI{10}{GHz} (Newport 818-BB-51A), $\text{PD}_{\text{r,fast}}$ to measure the fast part of the signal.

The output of the fast photodiode is externally amplified in two stages (Minicircuits ZX60-123LN-S+ and ZX60-02203+) before being mixed down and sent to the input of the LIA, which outputs a demodulated signal proportional to the component of the reflected light at the angular frequency $\omega_{\text{RF}}$ of the phase modulation frequency.

\section{Optical properties of the FFPC}

\subsection{Scan of the cavity resonance}

A key measurement done to characterize an optical cavity is to record the reflected and transmitted light, while continuously scanning the detuning between the laser and the cavity.
The FFPC's short length results in a large cavity linewidth (\si{\GHz}) and an even larger free spectral range (\SI{10}{\THz}).
Therefore, the best way to scan the detuning between the laser and cavity across a wide range of values -- on the order of a few cavity linewidths to a few free spectral ranges -- is to vary the length of the cavity.
This range of detuning corresponds to a change in cavity length on the order of \SI{100}{\pico\m} to \SI{3}{\um}, which is readily achieved using the piezoelectric positioner controlling the length of the FFPC.
Scanning the cavity length allows to perform such detuning scans rapidly compared to scanning the laser frequency, reducing the effects of mechanical drifts or low-frequency noise on the measurement.

In order to achieve the desired cavity scan, an S-shaped voltage ramp is generated with a function generator and used to drive the $z$-direction of the positioner controlling the cavity length.
The ramp consists of a constant acceleration phase, followed by a constant velocity phase, followed by a constant deceleration phase.
A reverse ramp is then applied to return the cavity length to its initial value.
The constant acceleration and deceleration phases prevent slip-stick motion of a stepper positioner along its guiding rod during the scan, improving the repeatability of the scans.

\Cref{fig:setup} (b) shows the cavity reflection signal $\text{PD}_r$ measured around a fundamental cavity resonance.
The scan in cavity length is expressed along the abscissa in terms of frequency, using the phase-modulation sidebands generated by the EOM at $\omega_\mathrm{RF}/(2\pi)=\SI{9.03}{\giga\hertz}$ as a reference.
The data is fit with the sum of three pseudo-Voigt-based lineshapes to extract the cavity linewidth $\kappa/(2\pi)=\SI{0.97}{\giga\hertz}$.

\Cref{fig:setup} (c) shows the cavity error signal used to stabilize the cavity.
This signal is generated using a modified PDH scheme, inspired by refs.~\onlinecite{KashkanovaOptomechanicsSuperfluidHelium2017,ShkarinQuantumOptomechanicsSuperfluid2018}.
The input light, whose phase is modulated at $\omega_{\mathrm{RF}}$, results in reflected light from the cavity, whose amplitude is modulated at the same frequency.
This signal is detected by the fast photodiode and demodulated by the mixer and LIA, yielding the cavity error signal \cite{BlackintroductionPoundDrever2000}.
This signal is used as the process variable in a proportional-integral-derivative (PID) feedback controller regulating the fiber cavity length via the fast piezoelectric ring.
The PID is realized using a RedPitaya FPGA, which is controlled by the PyRPL Python module \cite{NeuhausPyRPLPythonRed2017}.
The PID input includes a digital filter, which is designed in the Infinite Impulse Response (IIR) module, to prevent the feedback signal from driving system resonances, thus maximizing the bandwidth of the lock (see Section IV).
The detuning signal matches the expected PDH response, as shown by a fit \cite{BlackintroductionPoundDrever2000}, which uses parameters extracted from a fit of the cavity reflection signal along with two fit parameters: a scaling factor and a shift in the central detuning.
We attribute this shift to delay in the electronics.

\subsection{Characterization of the cavity finesse}

The cavity finesse is estimated using $\mathcal{F} = \pi c/(\kappa L_c) $, where $c$ is the speed of light and $L_c$ is the cavity length extracted using a white-light interferometry technique similar to the one introduced in ref.~\onlinecite{JiangHighresolutioninterrogationtechnique2008}. 
The details of this technique will be the subject of an upcoming paper.

\begin{figure}[!t]
	\centering\includegraphics{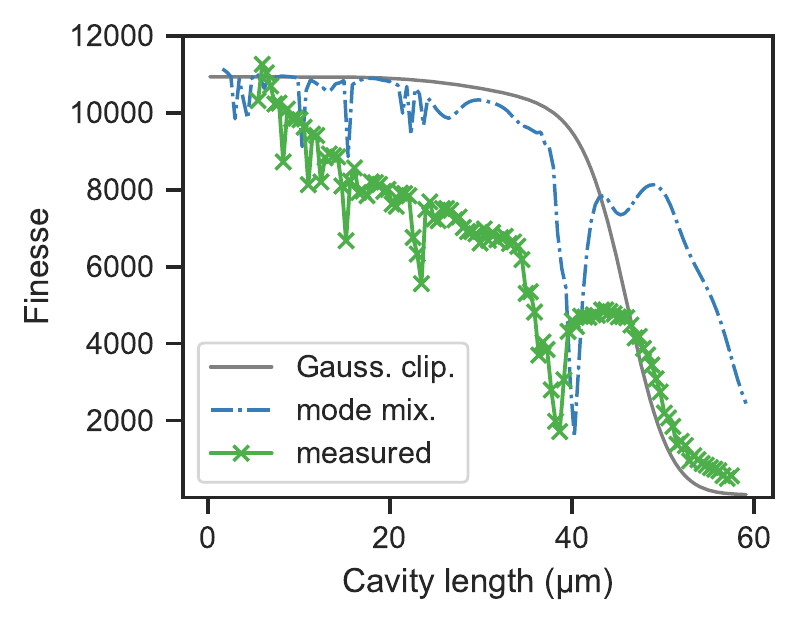}
	\caption{
		Cavity finesse (green) as a function of cavity length showing the large range over which we can tune our cavity. 
		The blue and grey curves are calculated with a mode-mixing simulation and a Gaussian clipping model, respectively.	
	}
	\label{fig:finesse}
\end{figure}

\Cref{fig:finesse} shows the results of the finesse measurement in green as a function of the length of the fiber cavity.
The finesse is recorded starting from a very short mirror separation, at which the optical fibers are almost in contact and it reaches a maximum of $10^4$, ending at a cavity length of \SI{60}{\micro\meter} where the finesse has deteriorated such that the reflected signal is no longer observable.

For comparison, a numerical simulation of the cavity resonance spectrum and of the associated losses has been performed following the formalism introduced in \cite{BenedikterTransversemodecouplingdiffraction2015} and using the actual shapes of the fiber-mirrors measured by a profilometer.
This simulation is carried out over the full range of measured cavity lengths and the result is plotted in blue.

A second comparison is made to the finesse calculated using a standard Gaussian clipping losses model \cite{HungerfiberFabryPerot2010}, plotted in grey as a function of the fiber-cavity length.
The simulated and calculated finesse were both rescaled to match the measured finesse for the shortest cavity lengths.

Three components of the evolution of finesse can be distinguished in the measured data plotted in \cref{fig:finesse}.
The first is a rapid decrease in finesse as the cavity length approaches \SI{50}{\micro\meter}.
This behavior matches the Gaussian clipping model and is due to the spot-size of the fundamental cavity mode approaching the aperature diameter $D_\mathrm{sph}$ of the mirrors.
The second effect is the presence of sharp dips in finesse as a function of cavity length, which is reproduced by the mode-mixing simulation.
The dips occur at cavity lengths where higher order modes of longitudinal order $q$ become resonant with the fundamental mode of longitudinal order $q+1$.

Finally, the third effect is the overall decrease in finesse in the measured data as the cavity gets longer.
We attribute this monotonic decrease to a misalignment of the cavity as its length increases.
Because corrections to the cavity's lateral alignment were impractical during this measurement, \cref{fig:finesse} should be considered a lower bound on the finesse at each cavity length.
If desired, the apparatus makes it possible to maximize the finesse for any given cavity length.

These measurements show that our FFPC can be tuned and operated over a large range of cavity lengths while maintaining good alignment.
This capability allows for integration with a variety of different systems in the middle of the cavity and for the exploration of different parameter regimes.
Note that parameters such as the maximal length at which our cavity can be operated are strongly influenced by the mirror geometry, which can be modified by tuning laser ablation parameters \cite{RuelleOptimizedsingleshotlaser2019}.

\section{Passive stabilization}

The detrimental effects of mechanical noise on the stability of the optical cavity are most obvious when monitoring the cavity resonance.
The FFPC is extremely sensitive to displacement of the fiber mirrors along its optical axis, which induces variations in the cavity length and thus in the detuning between the cavity and the fixed laser wavelength.
Lateral motion of the fiber mirrors contributes as a second-order effect, causing misalignment of the cavity, which in turn modulates the reflected and transmitted signals.
Such lateral motion can also result in small variations in the cavity length due to the shape of the mirrors.

\begin{figure*}[!t]
	\centering\includegraphics[width=17cm]{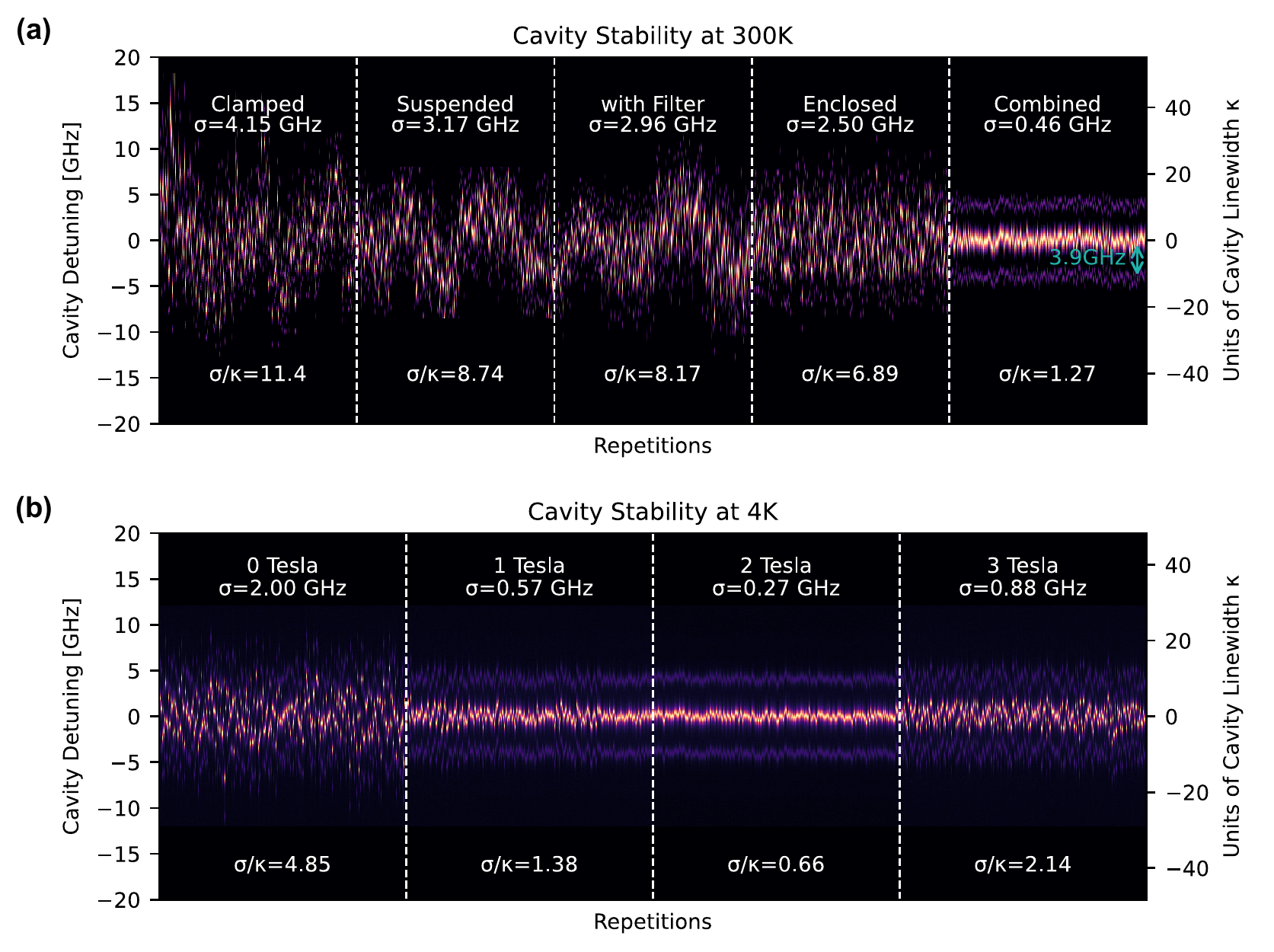}
	\caption{Repeated measurements of the light
          transmitted through the cavity as a function of
          its length both at room temperature (a) and low
          temperature (b), showing the impact of different system
          configurations on the stability of the cavity.  }
	\label{fig:system_stability}
\end{figure*}

The high finesse of the FFPC of around $10^4$ results in a cavity resonance, whose width corresponds to a cavity displacement of
\SI{50}{\pm}.
Such a small linewidth makes the cavity a highly nonlinear transducer of any changes in cavity length on the order of or larger than this width \cite{ReinhardtSimpledelaylimitedsideband2017}.
As a result, both the characterization and the stabilization via feedback of the cavity displacement noise is technically challenging.
Therefore, we first use passive stabilization in order to reduce displacement noise down to the level of a cavity linewidth or less, at which point it can be actively stabilized using the PDH technique.
Passive stabilization is achieved by shielding the FFPC from mechanical noise sources and by damping its mechanical vibrations.  

In order to get a rough estimate of the cavity stability, we perform repeated measurements of the light transmitted through the cavity as a function of its length around a fundamental resonance.
Assuming that run-to-run variations in the position of the resonance peak are dominated by variations in the length of the cavity and not by the repeatability of the piezoelectric displacement scan, this measurement provides a measure of the FFPC's mechanical stability.

\Cref{fig:system_stability} shows two sets of these measurements, one carried out at room temperature (a) and one at \SI{4}{\kelvin} (b).
Each vertical line represents a cavity scan, and each cavity scan reveals a main peak and two less intense phase-modulation sidebands split by \SI{3.9}{\giga\hertz} from the main peak.
A measure of the displacement noise is provided by the standard deviation of the main peak position both in units of frequency and in terms of the cavity linewidth.
The cavity linewidth is estimated to be \SI{0.36}{\giga\hertz} at room temperature and \SI{0.41}{\giga\hertz} at low temperature, using the average of Lorentzian fits to each resonance peak.
The cavity length, as determined by white light interferometry (see above), is $L_{c}=\SI{25}{\micro\meter}$ at room temperature and $L_{c}=\SI{24}{\micro\meter}$ at \SI{4}{\kelvin}.
At room temperature, a number of different experimental conditions, e.g.\ different combinations of suspension, electrical filtering, and pressure, were studied.
Those shown in \cref{fig:system_stability} (a) cover a selection of conditions, that shed light on what is required to best passively isolate the system from external noise.
The measurements shown in (b) were taken at \SI{4}{\kelvin}, here we explore the effect of an applied magnetic field on the cavity stability.
Since this magnetic field is generated by a superconducting magnet that requires cryogenic temperatures, this measurement could only be performed at \SI{4}{\kelvin}.

\begin{figure*}
	\centering\includegraphics[width=15cm]{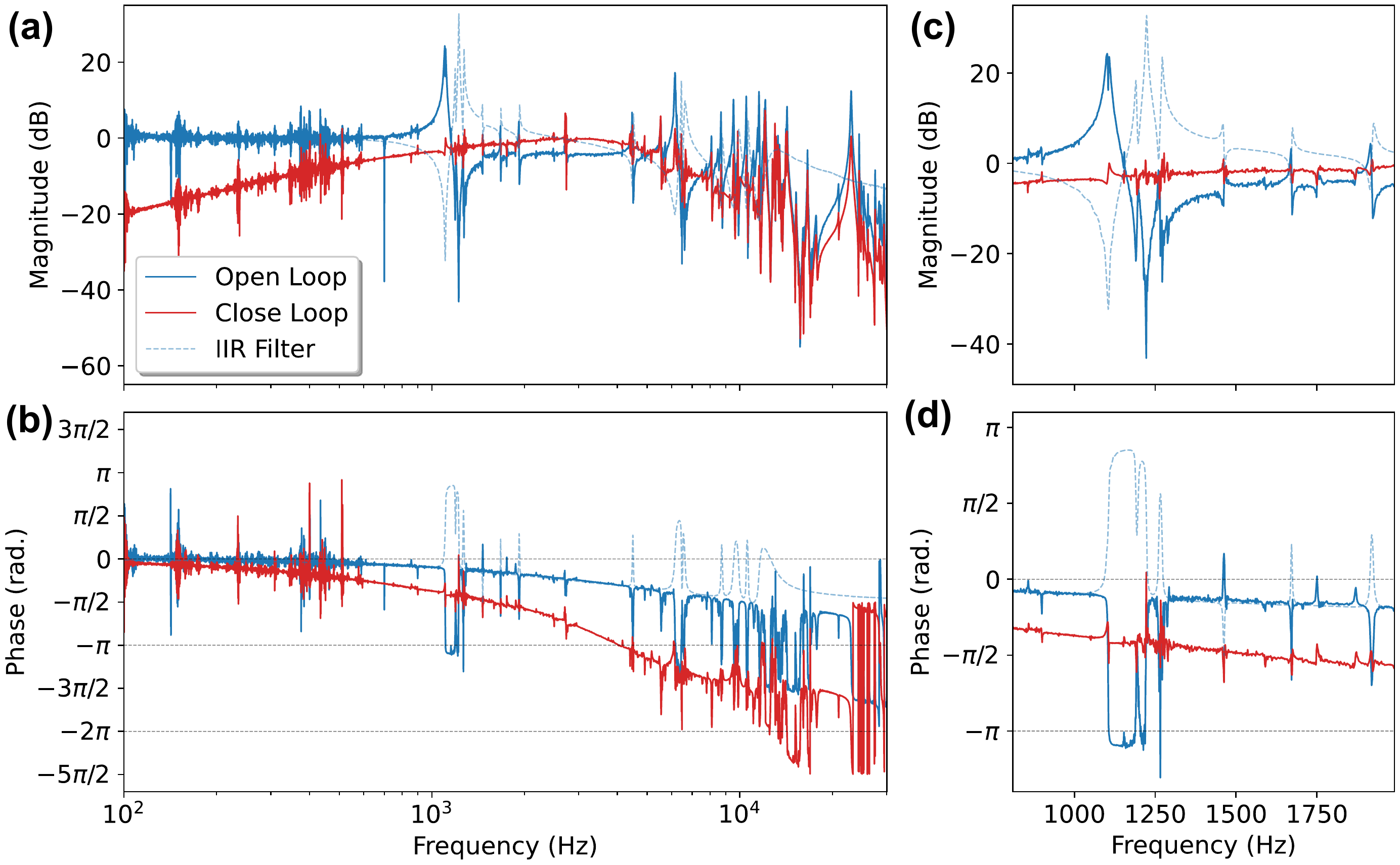}
	\caption{
		Feedback loop transfer function (amplitude (a) and phase (b)) of the cavity lock: in blue, the system response in the quasi-open loop configuration.
		It is possible to identify many resonances in the system, mostly due to the piezo motors and to the cantilever modes of the fiber mirror.
		The dotted line is the IIR digital filter designed to suppress most of the system resonances below \SI{20}{\kHz} (limited by the computational power of the FPGA).
		The red line is the system response when the IIR filter is enabled.
		(c) and (d) are a zoom around the most problematic region around \SI{1}{\kHz}.
	}
	\label{fig:system_transfer_function}
\end{figure*}

The first room-temperature configuration, labeled \textquote{Clamped} in \cref{fig:system_stability} (a), serves as a baseline for comparison.
In this configuration, the cryogenic probe is outside of the cryostat, the titanium frame containing the FFPC is rigidly clamped to the probe, the electrical filters are turned off, and the cavity is open to the environment.
The second configuration, labeled \textquote{Suspended}, differs from the first only in that the titanium frame is suspended using springs, resulting in a slight improvement.
The third configuration, labeled \textquote{with Filter}, differs from the first only in that the electronic filters are activated to reduce the noise on the voltage driving the piezoelectric positioners.
The specified output noise of the amplifiers (Attocube ANC300) is less than \SI{5}{\milli\volt}, which corresponds to a displacement of the fiber mirror by 5 times the cavity linewidth at room temperature and 0.5 times the cavity linewidth at \SI{4}{\kelvin}.
The filter used is a third-order low-pass filter at \SI{10}{\Hz}.
Note that despite the $z$-axis being the biggest source of noise for the cavity, all the axes (including the positioners in the sample block), need to be filtered or physically disconnected from the system for optimal results.
The fourth configuration, labeled \textquote{Enclosed}, differs from the first by including both the suspension of the titanium frame via springs and its enclosure inside the vacuum can.
This configuration appears to suppress slow modulations of the cavity length, which are likely due to air currents.
Whether or not the system is in vacuum does not notably affect the mechanical noise.
Finally, the fifth configuration, labeled \textquote{Combined}, includes the spring suspension, the electronic filters, and the enclosure in the vacuum can.
The combination of these measures results in a passive stability close to the cavity linewidth, thus bringing the system into a regime, in which active PDH stabilization can be employed.

\begin{figure*} 
	\centering\includegraphics[width=15cm]{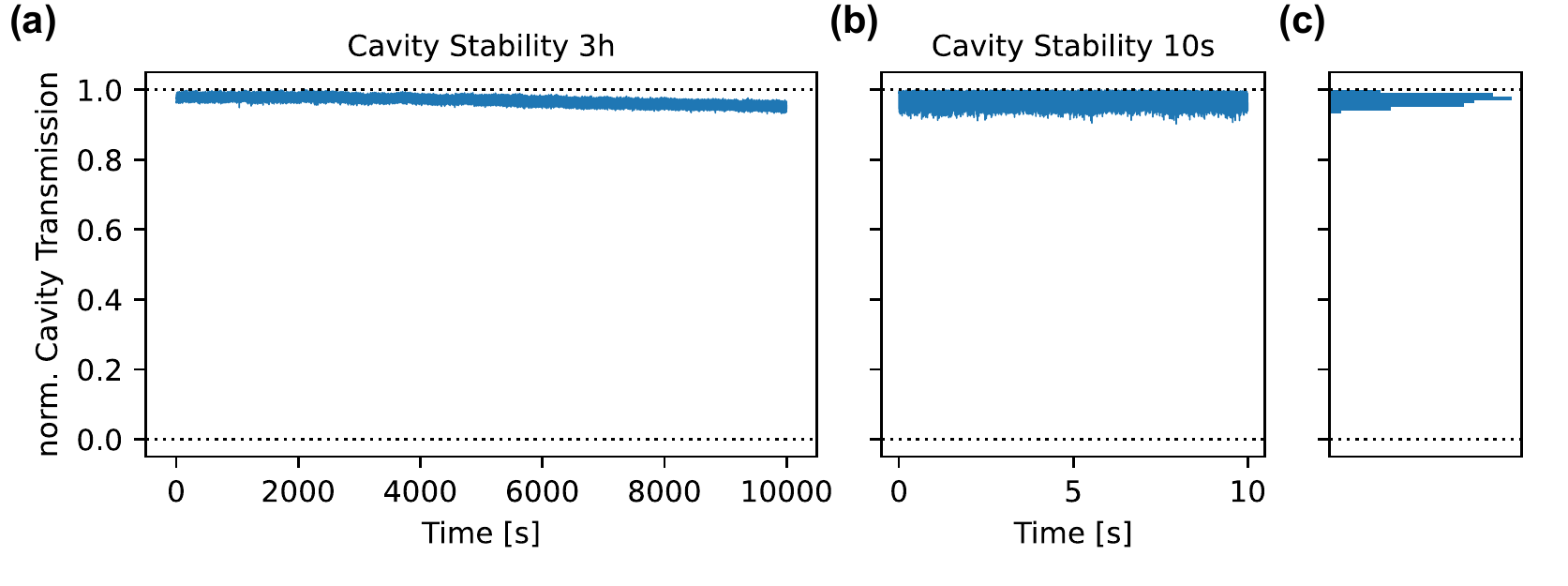}
	\caption{
		Long term stability of the cavity transmission signal while the cavity is locked.
		(a) shows a stability measurement over 3 hours, the dashed line at 1.0 showing the maximum cavity transmission; we attribute the drift towards lower values of the cavity transmission signal to power instability of the optical measurement setup, not to an instability of the cavity itself.
		(b) shows a shorter term version of the same measurement, revealing a clearly visible maximum value when the cavity is on resonance, again shown by a dashed line at 1.0.
		(c) shows a histogram of the taken measurement points in (b).
	}
	\label{fig:lock_stability}
\end{figure*}

For all of the measurements carried out at \SI{4}{\kelvin}, the cryogenic probe is inserted in the cryostat and the FFPC system is suspended and in vacuum.
The electronic filters do not have a noticeable effect under these conditions, probably because of the 10-fold reduction in the piezoelectric coefficient of the positioners from room temperature to \SI{4}{\kelvin}.
In the absence of applied magnetic field, the system shows a reduced stability compared to the same configuration, labeled \textquote{Combined}, at room temperature.
This deterioration may be caused by bubbling of the liquid helium or nitrogen that surround the probe or by a temperature-dependent change in the mechanical properties of the probe materials.
The application of a magnetic field, however, dampens vibrational noise via eddy-currents induced in the suspended conducting frame of the FFPC.
At \SI{2}{\tesla} of applied field, this effect brings the stability to better than a cavity linewidth and thus into a regime in which the cavity can be actively stabilized.
A subsequent increase in vibrational noise observed at \SI{3}{\tesla} may be the result of a residual magnetization in some part of the suspended system, causing it to deflect in the magnetic field and come in contact with the vacuum can.

\section{Active stabilization}
Once a sufficient passive stability has been achieved, the cavity length is locked with respect to the fixed laser wavelength via active feedback using a modified PDH scheme that acts on the fast piezoelectric ring, which directly modulates the cavity length.
In order to characterize the transfer function of the system and optimize the performance of this feedback loop, we performed a response analysis of the system's transfer function, as shown in \cref{fig:system_transfer_function}.
The vector network analyzer module of the PyRPL GUI is used to drive the piezoelectric ring at a frequency that is ramped between \SI{100}{\hertz} and \SI{30}{\kilo\hertz}, while recording the magnitude and phase of the cavity's PDH response.
This measurement is first performed with the system in a quasi open-loop configuration, i.e.\ the feedback signal sent to the ring piezo is low-pass filtered at \SI{10}{\Hz}, so as to only cancel low-frequency drifts of the cavity resonance.
The magnitude and phase of the system response detected in the PDH error signal, shown in blue in \cref{fig:system_transfer_function} (a) and (b), are measures of the system's mechanical resonances.
The first group between \SI{800}{\Hz} and \SI{2}{\kHz}, shown in the zoom in \cref{fig:system_transfer_function} (c), are likely resonances of the positioners.
Changes in alignment, via motion of the positioners, alters their amplitude and slightly changes their frequency.
The next prominent resonances appear above \SI{6}{\kHz}, and are at least in part due to the cantilever modes of the fiber mirrors.
Their frequencies match the expected resonance frequencies for fibers protruding by about \SI{4}{\mm} from their support ferrules.
The frequencies of all of these resonances slowly vary in time at room temperature and stabilize at \SI{4}{\kelvin}, pointing to a contribution from thermal drift.
An overall low-pass filter behavior, partially masked by the various resonances, is the result of the response of the ring piezo.

The system's mechanical resonances hinder the active stabilization of the cavity length by reducing the bandwidth and the maximum gain of the feedback loop.
Using a digital filter to suppress the components of the feedback signal at the frequencies of these resonances prevents spurious excitation of the system.
The fact that the resonances drift in time and shift upon realignment requires the use of a reconfigurable filter to suppress their response successfully.
Therefore, we employ an infinite impulse response (IIR) filter to cancel the resonances in the feedback signal before sending it as a drive to piezoelectric ring.
The IIR filter is designed in the IIR module of the PyRPL GUI by manually placing zeros and poles, in order to obtain a filter that is as close to the inverse of the system transfer function as possible, shown as the dotted blue line in \cref{fig:system_transfer_function}.
A maximum of $13$ pairs of zeros and poles can be used, limited only by the computational power of the FPGA, together with an additional real pole that act as a low-pass filter to stabilize the filter response.

The same response analysis is then performed in closed-loop configuration, with the feedback signal being sent to the piezoelectric ring after application of the IIR filter.
By virtue of the digital filter, the closed-loop response, shown in red in \cref{fig:system_transfer_function}, heavily attenuates the system's mechanical resonances, giving the stabilization loop a bandwidth of $4$ to \SI{6}{\kilo\hertz} at room temperature (corresponding to the frequency at which the phase crosses $-\pi$ \cite{BechhoeferFeedbackphysiciststutorial2005,BlackintroductionPoundDrever2000}) and of \SI{10}{\kilo\hertz} at \SI{4}{\K}.

The stability achieved by the system is demonstrated by recording the FFPC's transmission signal over an extended period of time, as shown in \cref{fig:lock_stability} (a).
This representative plot shows the transmission signal with the cavity locked over a period of \SI{3}{\hour}, although such stability can be sustained for a much longer periods without the need of a re-locking system.
The slow decrease in signal in the plot are the result of a drifts in the input laser power, which is not stabilized in this measurement.
A plot of a \SI{10}{\second} segment of the locked transmission signal, shown in \cref{fig:lock_stability} (b), reveals -- as expected -- that the transmission is limited to a maximum value with the cavity on resonance.
The cavity detuning stays within a few percent of this maximum which, given the cavity linewidth of \SI{50}{\pm}, corresponds to a stability better than a few picometers.

\section{Conclusions}


\definecolor{header}{RGB}{200,200,200}
\definecolor{gray}{RGB}{235,235,235}
\definecolor{white}{RGB}{255,255,255}
\renewcommand{\arraystretch}{1.5}
\begin{table*}[!htb]
\centering
\begin{threeparttable}[c]
	
\begin{tabularx}{\textwidth}{|X|Y|Y|Y|Y|Y|Y|}\hline
& \textbf{Zhong} \cite{Zhongmillikelvinallfibercavity2017}  & \textbf{Salz} \cite{SalzCryogenicplatformcoupling2020} & \textbf{Janitz} \cite{JanitzHighmechanicalbandwidth2017} & \textbf{Fontana} \cite{Fontanamechanicallystabletunable2021a} & \textbf{Vadia} \cite{VadiaOpenCavityClosedCycleCryostat2021} & \textbf{This work}  \\ \hline
\textbf{Cavity Type}  & Fiber-Fiber & Fiber-Mirror & Fiber-Mirror & Fiber-Mirror & Fiber-Mirror & Fiber-Fiber \\ \hline
\textbf{Purpose} & MIM Optomechanics & Coupling to \linebreak SiVs & Coupling to\linebreak NVs & Coupling to\linebreak NVs & QED (WSe\textsubscript{2}) & MIM Optomechanics \\ \hline
\textbf{Temperature\newline(cryostat) }& \SI{400}{\milli\kelvin}\linebreak(dil. fridge) & \SI{4}{\kelvin}\linebreak(dil. fridge) & -\linebreak(closed cycle) & \SI{11}{\kelvin}\linebreak(closed cycle) & \SI{7}{\kelvin}\linebreak(closed cycle) & \SI{4}{\kelvin} (bath\linebreak cryostat) \\ \hline
\textbf{Cavity Length} & \SI{23.7}{\micro\meter} & \SI{1.6}{\micro\meter}-\SI{45}{\micro\meter} & \SI{15}{\micro\meter} & \SI{18.7}{\micro\meter} & \SI{5}{\micro\meter} & \SI{1}{\micro\meter}-\SI{50}{\micro\meter} \\ \hline
\textbf{z tunability} & stepper motor + tube piezo & nanopositioner + shear piezo & shear piezo + temperature & tripod + shear piezo & nanopositioner + piezo element & nanopositioner + ring piezo \\ \hline
\textbf{Alignment} & fully tunable stepper motor system & xyz nanopositioner & fiber aligned during gluing & tripod + positioner stage & xyz nanopositioner & xyz nanopositioner + tip/tilt mount \\ \hline
\textbf{Finesse} & 50 -- 120 & 1k \tnote{a} & 21k  & 16k & 110 & 11k \\  \hline
\textbf{Passive\newline Stability} & -\linebreak- & \SI{290}{\pico\meter} (RT)\linebreak \SI{260}{\pico\meter} (\SI{4}{\K}) & -\linebreak - & -\linebreak \SI{16}{\pico\meter} (\SI{11}{\K}) & \SI{32}{\pico\meter} (RT)\linebreak \SI{117}{\pico\meter} (\SI{7}{\K}) & \SI{35}{\pico\meter} (RT)\linebreak \SI{20}{\pico\meter} (\SI{4}{\K})  \\ \hline
\textbf{Locked\linebreak Stability} & -\linebreak - & \SI{60}{\pico\meter} (RT)\newline \SI{90}{\pico\meter} (\SI{4}{\K}) & -\linebreak - & -\linebreak - & -\linebreak \SI{89}{\pico\meter} (\SI{7}{\K}) & \SI{3}{\pico\meter} (RT)\linebreak \SI{5}{\pico\meter} (\SI{4}{\K}) \\ \hline
\textbf{Lock\linebreak Bandwidth} & -\linebreak - & \SI{800}{\hertz} (RT)\linebreak \SI{300}{\hertz} (\SI{4}{\K}) & \SI{44}{\kilo\hertz} (RT)\linebreak -\tnote{b} & \SI{10}{\kilo\hertz} (RT)\linebreak \SI{1}{\kilo\hertz} (\SI{11}{\K})\tnote{c} & -\linebreak \SI{50}{\hertz} (\SI{7}{\K})\tnote{d} & \SI{6}{\kilo\hertz} (RT)\linebreak \SI{10}{\kilo\hertz} (\SI{4}{\K}) \\ \hline
\end{tabularx}
\begin{tablenotes} 
\item [a] at the wavelength of the lock laser.
\item [b] no successful operation at LT.
\item [c] active lock limited to quiet period of the closed cycle cryostat.
\item [d] active lock has moderate impact at LT.
\end{tablenotes}
\caption{
	Overview of the different implementations of cryo-compatible FFPCs. 
	Comparing the performance and tunability of the system presented in this work with that of other published systems shows that it achieves a unique combination of a high finesse, high stability, and large tunability, enabled by the successful cavity lock at both room and cryogenic temperatures.
	SiV: silicon vacancy center (in diamond), NV: nitrogen vacancy center (in diamond), QED: quantum electrodynamics, WSe\textsubscript{2}: tungsten diselenide (excitonic transition).
}
\label{tab:system_comparison}
\end{threeparttable}
\end{table*}

Our experimental apparatus shows that it is possible to build a highly tunable cryogenic FFPC without sacrificing in key performance parameters, such as finesse or stability.  The apparatus maintains a high finesse over a large range of cavity lengths with values as high as $10^4$ for the shortest cavity lengths.  Measures to shield the system from vibrational noise yield a passive stability on the order of the cavity linewidth both at room temperature and \SI{4}{\kelvin}.
In this regime, active stabilization via a digitally filtered feedback loop results in a final stability of \SI{3}{\pico\meter} at room temperature and \SI{5}{\pico\meter} at \SI{4}{\kelvin}.  

\Cref{tab:system_comparison} shows the key properties of our apparatus along with those of other cryogenic FFPCs described in the literature.
In this comparison, the versatility and benefits of this system become clear.
In particular, the performance of our apparatus' active stabilization stands out.
To the best of our knowledge, all actively stabilized FFPCs either feature a low finesse cavity \cite{Zhongmillikelvinallfibercavity2017,VadiaOpenCavityClosedCycleCryostat2021}, do not operate at cryogenic
temperatures \cite{BrandstatterIntegratedfibermirrorion2013,SaavedraTunableFiberFabryPerot2021,RochauDynamicalBackactionUltrahighFinesse2021}, are not tunable in-situ \cite{SaavedraTunableFiberFabryPerot2021,JanitzHighmechanicalbandwidth2017}, or are unable to maintain a lock for extended periods of time at cryogenic temperatures \cite{JanitzHighmechanicalbandwidth2017,Fontanamechanicallystabletunable2021a}.
In particular, the active stability of many of these systems is limited by spurious mechanical resonances, whose deleterious effects could be reduced by the IIR filtering scheme presented in section IV.

We note that a pair of simple improvements to the current apparatus should result in further gains in performance.
First, the finesse of the FFPC is currently limited by the fiber-mirror coating, which is designed for a finesse of $1.5 \times 10^4$.
The fabrication of new fiber-mirrors with an ultra-low-loss coating should lead to a reduction in the cavity linewidth from $\kappa/(2\pi) \approx 500$ to $\SI{50}{\MHz}$ and, consequently, to an increase of the finesse by an order of magnitude.
Second, the bandwidth of the active stabilization is partially limited by the low-frequency cantilever mode of the overhanging fiber-mirrors.
By slightly modifying the design of the fiber supports, this overhang can be drastically reduced, leading to improved active stabilization.
Given the current performance of the stabilization, we expect to be able to reliably lock a cavity with a finesse of around $10^5$, while maintaining all other advantages of the system.

Because of its high sensitivity and its versatility, the experimental platform introduced here is suitable for studying a broad range of nanomechanical resonators, including silicon carbide nanowires \cite{FoglianoMappingCavityOptomechanical2021}, nanowires with embedded emitters \cite{MullerCouplingepitaxialquantum2009}, carbon nanotubes \cite{StapfnerCavityenhancedopticaldetection2013}, magnetic 2D materials \cite{SiskinsMagneticelectronicphase2020}, or functionalized membranes \cite{HalgMembraneBasedScanningForce2021}.
Initial experiments in the current system are underway with a micromechanical hBN drum resonator in the middle of the FFPC.
In this membrane-in-the-middle configuration, the membrane can be freely positioned within the cavity and various optomechanical measurements have been carried out and will be reported elsewhere.


\begin{acknowledgments}
We thank Sascha Martin and his team in the machine shop of the Physics Department at the University of Basel for help building the apparatus.
We acknowledge the support of the Canton Aargau and the Swiss National Science Foundation (SNSF) under Ambizione Grant No. PZ00P2-161284/1, Project Grant No. 200020-178863, and via the NCCR Quantum Science and Technology (QSIT).
\end{acknowledgments}

%
%

%

\end{document}